\RequirePackage{ifpdf}
\ifpdf 
\documentclass[pdftex]{sigma}
\else
\documentclass{sigma}
\fi

\begin{document}
\allowdisplaybreaks

\renewcommand{\PaperNumber}{045}

\FirstPageHeading

\ShortArticleName{Electroweak Interaction Model with an
Undegenerate Double Symmetry}

\ArticleName{Electroweak Interaction Model\\ with an Undegenerate
Double Symmetry}

\Author{Leonid M. SLAD} \AuthorNameForHeading{L.M. Slad}

\Address{D.V. Skobeltsyn Institute of Nuclear Physics, Moscow
State University, Moscow, 119899 Russia}
\Email{\href{mailto:slad@theory.sinp.msu.ru}{slad@theory.sinp.msu.ru}}

\ArticleDates{Received December 12, 2005, in f\/inal form March
31, 2006; Published online April 20, 2006}

\Abstract{The initial $P$-invariance of the electroweak
interaction Lagrangian together with the low-energy results of the
Weinberg--Salam model is provided by a local secondary symmetry.
Among the transformation parameters of this symmetry there are
both scalars, and pseudo-scalars with respect to the orthochronous
Lorentz group. Such symmetry does admissible existence of a light
(massless) axial gauge boson and its possible nonuniversal
interaction with the leptons of various types.}


\Classification{81T10; 81R05; 81V10}

\section{Introduction}
The basic point of this report is a logic precondition for
possible existence of a light (massless) axial gauge boson which
may interact with the electronic neutrino. This precondition is
based on a symmetry approach, which has been formulated rather
recently, though separate examples of its realization are known
for a long time. Prior to passing to the basic theme of the
report, we consider it necessary to give some characteristics of
the double symmetry concept used by us, to list the existing
undegenerate double symmetries in field theories, and to formulate
the principal positions of the initially $P$-invariant model of
electroweak interactions.

\section{Some characteristics of the double symmetry}

The concepts of the secondary and double symmetry is proposed by
us \cite{Slad00} as some generalization of already existing
approaches to the field theories construction.

The double symmetry group ${\cal G}_{T}$ is always the semidirect
product ${\cal G}_{T} = H_{T} \circ G$ of the subgroup~$G$, which
is the global primary symmetry group, and the invariant subgroup
$H_{T}$, which is the global or local secondary symmetry group.

The secondary symmetry group $H_{T}$ is generated by
transformations, whose parameters $\theta =  \{\theta_{a}\}$ are
vectors of the space of a set beforehand representation $T$ of the
primary symmetry group $G$. Let us emphasize that in the general
group theory, the situation when the parameters of one group
transform as a nontrivial representation of another one is not
discussed at all.

We realize both a representation $S$ of the primary symmetry group
$G$, and the secondary symmetry transformations in the form
\begin{gather}\label{equation1}
\Psi '(x) = \exp (-iD^{a}\theta_{a})\Psi (x)
\end{gather}
in the same space of field vectors $\Psi(x)$. Generally the
secondary symmetry transformations~\eqref{equation1} connect among
themselves both the vectors of the same irreducible
representation, and the vectors of various irreducible
representations belonging to the representation $S$ of group $G$.

The secondary symmetry transformations do not violate the primary
symmetry. It means that the operators $D^{a}$ in
equation~\eqref{equation1} should be the $T$-operators of the
group $G$, i.e.
\begin{gather*}
D^{a} = S^{-1}(g) D^{b} S(g) [T(g)]_{b}{}^{a}.
\end{gather*}

If the secondary symmetry is produced by the adjoint
representation of the group $G$ and the operators $D^{a}$ in
equation~\eqref{equation1} coincide with the group generators,
then obviously the group $H_{T}$ is locally isomorphic to the
group $G$. In such a case we say that the double symmetry is
degenerate.

\section{Existing undegenerate double symmetries in field theories}

{\bf A. The ${\boldsymbol \sigma}$-model symmetry.} The
$\sigma$-model symmetry of Gell-Mann and Levy \cite{Gell-Mann} is
nothing but an undegenerate double symmetry. Transformations in
the $\sigma$-model connect the pseudoscalar $\pi$-meson and the
scalar $\sigma$-meson. Their infinitesimal form is
\begin{gather}\label{equation3}
{\boldsymbol \pi}' = {\boldsymbol \pi} + i {\boldsymbol \theta}
\sigma ,
\\
\label{equation4} \sigma' = \sigma - i {\boldsymbol \theta}
{\boldsymbol \pi} .
\end{gather}
Gell-Mann and Levy obviously note that in transformations
\eqref{equation3} and \eqref{equation4} the parameter
${\boldsymbol \theta}$ is a~pseudoscalar with respect to the
orthochronous Lorentz group $L^{\uparrow}$. Due to this, these
transformations do not violate the spatial reflection symmetry P.

In the $\sigma$-model, the primary symmetry group $G$ is $SU(2)
\otimes L^{\uparrow}$. The secondary symmetry is produced by the
representation $T$ = ({\it isotriplet, pseudoscalar}) of $G$. Its
group is $H_{T} = SU(2)_{L} \otimes SU(2)_{R}$. The parameters of
one of the groups $SU(2)$ are given by the sum of the space scalar
and pseudoscalar, and the parameters of the other group are given
by their difference. It is necessary specially to note that,
unlike Gell-Mann and Levy, nobody except us \cite{Slad00} said
anything about the $P$-properties of parameters of the chiral
symmetry group $SU(2)_{L} \otimes SU(2)_{R}$.

{\bf B. Supersymmetry.} Supersymmetry, both in the $x$-space and
the superspace, can be considered as a secondary symmetry produced
by the bispinor representation (by a direct sum of two
nonequivalent irreducible spinor representations) of the proper
Lorentz group $L^{\uparrow}_{+}$, which plays the role of the
primary symmetry group. The set of generating elements for
supersymmetry consist of transformations of the form
\eqref{equation1}, in which the parameters $\theta =
\{\theta_{a}\}$ belong to the bispinor representation space of the
group $L^{\uparrow}_{+}$. Therefore, supersymmetry transformations
connect bosonic and fermionic fields. This connection does not
break the statistics of the field states, iff the parameters
$\theta_{a}$ are anticommutating elements of the Grassmann
algebra.

{\bf C. The Poincar\`{e} group $\cal{P}$ as a double symmetry
group.} It is well-known that the Poincar\`{e} group is the
semidirect product ${\cal P} = T_{4} \circ L^{\uparrow}$ of the
subgroup of orthochronous Lorentz transformations $L^{\uparrow}$
and the invariant subgroup of four-dimensional translations
$T_{4}$. Therefore, the translation group can be considered as a
secondary symmetry group produced by the polar four-vector
representation of the orthochronous Lorentz group, and thus,
$L^{\uparrow}$ is the primary symmetry group. The corresponding
secondary symmetry (translation) transformations have the form
\begin{gather}\label{equation5}
\Psi '(x) = \exp (-i D^{\mu} \theta_{\mu}) \Psi (x),
\end{gather}
where the parameters $\theta_{\mu}$ are four-vector translations,
and $D^{\mu}$ are {\it the differential operators}, namely
$D^{\mu} = i \partial^{\mu}$.

{\bf D. An infinite-component field theory with a double
symmetry.} We have received a number of physically interesting
results in the theory of ISFIR-class fields, which transform as
proper Lorentz group representations decomposable into an infinite
direct sum of finite-dimensional irreducible representations
\cite{Slad01,Slad02,Slad04,Slad05}.

The infinite number of arbitrary parameters in the
relativistically invariant Lagrangians of the free ISFIR-class
fields was the serious reason for that until recently, there were
no research of such field theory. This arbitrariness is eliminated
by the requirement for the theory to be invariant also under
secondary symmetry transformations of the form \eqref{equation5},
in which the parameters~$\theta_{\mu}$ are the components of the
polar or axial 4-vectors of the orthochronous Lorentz group, and
the operators $D^{\mu}$ have {\it the matrix realization}.

It is proved \cite{Slad05} that there are such theories of the
ISFIR-class fields with spontaneously broken secondary symmetry,
whose mass spectra are quite satisfactory from the standpoint of
particle physics. This result is especially important, because in
all previous attempts aimed at a consistent relativistic
description of particles with an infinite number of degrees of
freedom, the mass spectra were physically unacceptable, as they
had an accumulation point at zero. Let us note that such attempts
were undertaken in particular by the well-known physicists
V.L.~Ginzburg, I.E.~Tamm, H.~Yukawa, Yu.M.~Shirokov, and
M.A.~Markov (see the review paper \cite{Ginzburg}) and
mathematicians I.M.~Gelfand and A.M.~Yaglom (see the monograph
\cite{Gelfand}).

\section[Initially $P$-invariant electroweak model as a logical correction of
the standard left-right symmetric model]{Initially
$\boldsymbol{P}$-invariant electroweak model as a logical
correction\\ of the standard left-right symmetric model}

The Weinberg--Salam electroweak model with the gauge group
$SU(2)_{L} \otimes U(1)$ is initially asymmetric with respect to
the left-handed and right-handed spinors and, therefore,
noninvariant with respect to the spatial reflection $P$.

The standard left-right symmetric model of electroweak
interactions \cite{Pati,Mohapatra1,Mohapatra2,Senjanovic}
eliminating the mentioned asymmetry reproduced all results of the
Weinberg--Salam model in the region of existing energies. Possible
manifestations at high energies of the additional heavy bosons,
denoted as $W_{R}$ and $Z_{LR}$, are a subject of regular
experimental studies (see, for example,~\cite{Particle}). The
parameters of the gauge group $SU(2)_{L} \otimes SU(2)_{R} \otimes
U(1)$ of \cite{Pati,Mohapatra1,Mohapatra2,Senjanovic} are scalars
with respect to orthochronous Lorentz group $L^{\uparrow}$. Such
group transformations violate the $P$-symmetry. For example, they
transform a $P$-even fermionic state into a state with uncertain
$P$-parity. Thus, the left-right symmetry of
\cite{Pati,Mohapatra1,Mohapatra2,Senjanovic}
 does not entail the $P$-symmetry.

The concept of the left or right essence is logically strict only
concerning the irreducible spinor representations of the proper
Lorentz group $L^{\uparrow}_{+}$ and the currents constructed of
these spinors. It does not bear in itself any sense concerning the
integer spin fields and, hence, concerning the gauge and Higgs
fields. The mathematically exact concept, which bears a direct
relation to all elements of electroweak interactions, is the
$P$-transformation of any field, any current, etc. The
$P$-transformation converts, in particular, the left-handed spinor
into the right-handed one and vice versa. The $P$-symmetry entails
the left-right symmetry.

If we wish to have the initially $P$-invariant model of
electroweak interactions, it is necessary to construct it in the
spirit of Gell-Mann and Levy \cite{Gell-Mann}. Such construction
has been realized in the paper \cite{Slad00}, where initial
$P$-invariance and its observed violation has been ensured by the
local secondary symmetry produced by the representation $T$ =
({\it isotriplet, scalar}) $\oplus$ ({\it isotriplet,
pseudoscalar}) $\oplus$ ({\it isosinglet, scalar}) $\equiv$ $(1,s)
\oplus (1,p) \oplus (0,s)$ of the primary symmetry group $G =
SU(2) \otimes L^{\uparrow}$. The corresponding parameters of the
secondary symmetry transformations and the gauge fields are
denoted as ${\boldsymbol \theta}^{1s} = \{ \theta^{1s}_{j} \}$,
${\boldsymbol \theta}^{1p} = \{ \theta^{1p}_{j} \}$,
$\theta^{0s}$; ${\boldsymbol B}^{1V}_{\mu} = \{ B^{1V}_{j \mu}
\}$, ${\boldsymbol B}^{1A}_{\mu} = \{ B^{1A}_{j \mu} \}$,
$B^{0V}_{\mu}$ ($j = 1, 2, 3$; the indices $V$ and $A$ mean the
polar and axial 4-vectors, respectively).

As the gauge groups of both the standard left-right symmetric
model of electroweak interactions
\cite{Pati,Mohapatra1,Mohapatra2,Senjanovic}, and the primary
$P$-invariant model \cite{Slad00} are locally isomorphic, the
Lagrangian structure for these models and, hence, all their
numerical results are identical. But in the conceptual plan, the
initially $P$-invariant model gives several remarkable
consequences which are absent in the left-right symmetric one.

First, the Higgs field $\Phi$ should have both scalar, and
pseudoscalar components, namely, the isodoublets
$\phi^{\frac{1}{2}s}$ and $\phi^{\frac{1}{2}p}$. If the neutral
components of both of them take some nonzero vacuum expectation
values $\langle\phi^{\frac{1}{2}s}_{-1/2}\rangle = v_{s}$ and
$\langle \phi^{\frac{1}{2}p}_{-1/2}\rangle = v_{p}$ with $\arg
(v_{p}/v_{s}) \neq \pm \pi /2$, then the masses of the two
$W$-bosons, which are connected with the left and right charged
currents, are different, i.e.\ the $P$-symmetry is violated. Hence
the nature of $P$-symmetry violation is in that the physical
vacuum does not possess definite $P$-parity. Let us note that the
relation
\begin{gather*}
|v_{s}-v_{p}| \ll |v_{s}+v_{p}|
\end{gather*}
is a sufficient condition for reproducing all results of the
Weinberg--Salam model, which are obtained for already completed
experiments.

Second, the fields of all intermediate bosons constitute a
superposition of polar and axial 4-vectors, and these vectors have
equal weight in the fields of $W$-bosons. In other words, the
$P$-properties of the weak currents and the $W$- and $Z$-bosons
connected with the corresponding currents are similar.

The formulas, confirming these two statements, will be presented
in the following section.

\section[Initially $P$-invariant electroweak model with the light (massless)
axial gauge boson]{Initially $\boldsymbol{P}$-invariant
electroweak model\\ with the light (massless) axial gauge boson}

Look at the parameters of the local group of the secondary
symmetry, which are listed in the previous section. The isotriplet
parameters are both scalar, and pseudoscalar with respect to the
group $L^{\uparrow}$. But, among the isosinglet parameters, we
find only a scalar and we do not find a pseudoscalar. Such
inequality does not look sufficiently natural. We hope to analyse
in the future the physical consequences of its elimination, laying
here the grounds for such analysis.

So, we shall enlarge the list of parameters of the secondary
symmetry group, leading to the initially $P$-invariant model of
electroweak interactions, by one more parameter $\theta^{0p}$
which transforms as the representation ({\it isosinglet,
pseudoscalar}) of the primary symmetry group $G = SU(2) \otimes
L^{\uparrow}$. The corresponding field will be denoted as
$B^{0A}_{\mu}$.

In the fermionic sector we restrict ourselves to isodoublet
$\psi^{T} = (\nu_{e}, \; e)$ consisting of the fields of
electronic neutrino $\nu_{e}$ and electron $e$. Its
transformations of the secondary symmetry under considaration can
be written in the form
\begin{gather}\label{equation7}
\psi' = \exp \left( -\frac{i}{2}{\boldsymbol \tau}{\boldsymbol
\theta}^{1s} - \frac{i}{2} \gamma^{5} {\boldsymbol
\tau}{\boldsymbol \theta}^{1p} + \frac{i}{2} \theta^{0s} + \frac{i
c_{1}}{2} \gamma^{5} \theta^{0p} \right) \psi,
\end{gather}
where ${\boldsymbol \tau}$ are Pauli matrices and $c_{1} = 0$ or
$2\sqrt{2}$.

The global secondary symmetry transformations of the fields
${\boldsymbol B}^{1V}_{\mu}$ and ${\boldsymbol B}^{1A}_{\mu}$ are
\begin{gather*}
{\genfrac(){0cm}{0}{{\boldsymbol B}^{1V}_{\mu}} {{\boldsymbol
B}^{1A}_{\mu}}}' = \exp \left[ -i\left(
\begin{matrix}
{\boldsymbol t}&0\cr 0&{\boldsymbol t} \end{matrix} \right)
{\boldsymbol \theta}^{1s} - i\left(
\begin{matrix}
0&{\boldsymbol t}\cr {\boldsymbol t}&0 \end{matrix} \right)
{\boldsymbol \theta}^{1p} \right] {\genfrac(){0cm}{0}{{\boldsymbol
B}^{1V}_{\mu}} {{\boldsymbol B}^{1A}_{\mu}}},
\end{gather*}
where ${\boldsymbol t} = \{ t_{j} \}$ ($j = 1, 2, 3$) are the
generators of adjoint representation of the group $SU(2)$.

The doubly symmetric interaction Lagrangian of the gauge field and
leptonic field $\psi$ is fixed in the form
\begin{gather}\label{equation9}
{\cal L}_{\rm int} = -\frac{1}{2\sqrt{2}}\overline{\psi} \left( g
\gamma^{\mu} {\boldsymbol \tau} {\boldsymbol B}^{1V}_{\mu} + g
\gamma^{\mu} \gamma^{5} {\boldsymbol \tau} {\boldsymbol
B}^{1A}_{\mu} - g_{0} \gamma^{\mu} B^{0V}_{\mu} - c_{1} g_{A}
\gamma^{\mu} \gamma^{5} B^{0A}_{\mu} \right) \psi,
\end{gather}
where the coupling constants $g$, $g_{0}$ and $g_{A}$ are
positive.

We write down the transformations of the Higgs field $\Phi$ and
mass terms of the Lagrangian for the gauge fields
\begin{gather}\label{equation10}
{\genfrac(){0cm}{0}{\phi^{\frac{1}{2}s}}{\phi^{\frac{1}{2}p}}}' =
\exp \left[ -\frac{i}{2} \left( \begin{matrix} {\boldsymbol
\tau}&0\cr 0&{\boldsymbol \tau} \end{matrix} \right) {\boldsymbol
\theta}^{1s} - \frac{i}{2} \left( \begin{matrix} 0&{\boldsymbol
\tau}\cr {\boldsymbol \tau}&0 \end{matrix} \right) {\boldsymbol
\theta}^{1p} - \frac{i}{2} \theta^{0s} - \frac{i c_{2}}{2} \left(
\begin{matrix} 0&{1}\cr {1}&0 \end{matrix} \right) {\theta}^{0p}
\right]
{\genfrac(){0cm}{0}{\phi^{\frac{1}{2}s}}{\phi^{\frac{1}{2}p}}},
\\
\label{equation11} {\cal L}_{\rm mass} = |({\cal
M}^{\mu})^{\dagger} {\cal M}_{\mu}|,
\end{gather}
where
\begin{gather}
{\cal M}_{\mu} = - \frac{1}{2\sqrt{2}} \left[ g \left(
\begin{matrix} {\boldsymbol \tau}&0\cr 0&{\boldsymbol \tau}
\end{matrix} \right) {\boldsymbol B}^{1V}_{\mu} + g \left(
\begin{matrix} 0&{\boldsymbol \tau}\cr {\boldsymbol \tau}&0
\end{matrix} \right)
{\boldsymbol B}^{1A}_{\mu} + g_{0} {\boldsymbol B}^{0V}_{\mu}\right.\nonumber\\
\left.\phantom{{\cal M}_{\mu} =}{} + c_{2} g_{A } \left(
\begin{matrix} 0&{1}\cr {1}&0 \end{matrix} \right) {\boldsymbol
B}^{0A}_{\mu} \right] {\genfrac(){0cm}{0}{\langle
\phi^{\frac{1}{2}s}\rangle}{\langle
\phi^{\frac{1}{2}p}\rangle}},\label{equation12}
\end{gather}
$c_{2}$ being an arbitrary constant.

As it was pointed out by Weinberg \cite{Weinberg}, an essential
element in the analysis of any gauge theory  is the reduction of
the Lagrangian mass term to a diagonal form that means the
elimination of field oscillations, i.e.\ the elimination of
transitions of one fields into others.  As a~result of this
procedure, there appear the orthonormalized physical fields with
their masses and coupling constants. The physical fields represent
linear sums (orthogonal rotation) of the initial gauge fields. In
the Weinberg--Salam model, such rotation involves two neutral
gauge fields and is characterized by one angle which is the
Weinberg angle. In the model under consideration, an appropriate
orthogonal rotation involves four neutral gauge fields and it is
convenient not to introduce any angles.

As it follows from equations~\eqref{equation11} and
\eqref{equation12}, only two neutral gauge fields of the four ones
possess masses, and the other two fields are massless. One of the
massless gauge fields $A_{\mu}$ is the electromagnetic field and
is described by the polar four-vector
\begin{gather*}
A_{\mu} = \frac{1}{\sqrt{g^{2} +
g^{2}_{0}}}\big(g_{0}B^{1V}_{3\mu} + gB^{0V}_{\mu}\big).
\end{gather*}
The other massless gauge field $Y_{\mu}$ is described by the axial
four-vector
\begin{gather}\label{equation14}
Y_{\mu} = \frac{1}{\sqrt{g^{2} +
(c_{2}g_{A})^{2}}}\big(c_{2}g_{A}B^{1A}_{3\mu} +
gB^{0A}_{\mu}\big).
\end{gather}
The axial gauge field $Y_{\mu}$ can get some small mass due to
another Higgs field (besides the field~$\Phi$), or due to a
mechanism we yet know nothing about.

Only such situation seems interesting to us when the axial gauge
boson interacts with the electronic neutrino and there is an
energy region in which this interaction can in principle be
experimentally detected.

As to the interactions of the axial field $Y_{\mu}$ with other
fields, there are two essentially different scenarios.

In the first variant the condition $c_{2} \neq 0$ is satisfied and
consequently the field $B^{1A}_{3\mu}$ has a~nonzero projection on
the state $Y_{\mu}$. It means that the axial boson $Y$ interacts
with one or two components of each fermionic isodoublet and with
the massive gauge fields. The axial boson interacts with the
electronic neutrino, if $c_{1} \neq c_{2}$.

In the second variant the constant $c_{2}$ is zero. Therefore, the
axial gauge boson can interact only with fermions, and only with
those ones, whose transformations of the type \eqref{equation7}
really contain the parameter $\theta^{0p}$. In particular, it
interacts with the electronic isodoublet, if $c_{1} \neq 0$. So,
in this variant of the electroweak model, the universality of all
interactions of the charged or neutral leptons is not obligatory.

The fields of comparatively light $W^{(1)\pm}_{\mu}$,
$Z^{(1)}_{\mu}$ and heavy $W^{(2)\pm}_{\mu}$, $Z^{(2)}_{\mu}$
intermediate bosons and their masses are described by the
following relations
\begin{gather*}
W^{(1)\pm}_{\mu} = \frac{1}{2} \big[\big(B^{1V}_{1\mu} -
B^{1A}_{1\mu}\big) \mp i\big(B^{1V}_{2\mu} -
B^{1A}_{2\mu}\big)\big],\nonumber
\\
W^{(2)\pm}_{\mu} = \frac{1}{2}
\big[\big(B^{1V}_{1\mu} + B^{1A}_{1\mu}\big) \mp
i\big(B^{1V}_{2\mu} + B^{1A}_{2\mu}\big)\big],
\\
Z^{(1)}_{\mu} = \sqrt{\frac{g^{2}+g^{2}_{0}}{2g^{2}+g^{2}_{0}}}
\left[ \frac{g}{g^{2}+g^{2}_{0}} (gB^{1V}_{3\mu} -
g_{0}B^{0V}_{\mu}) - B^{1A}_{3\mu} \right] + {\cal O}(x) + {\cal
O}(g_{A}),\nonumber
\\
Z^{(2)}_{\mu} = \frac{1}{g^{2}+g^{2}_{0}} \big(gB^{1V}_{3\mu} -
g_{0}B^{0V}_{\mu} + gB^{1A}_{3\mu}\big) + {\cal O}(x) + {\cal
O}(g_{A}),\nonumber
\\
m^{2}_{W^{(1)}} = \frac{g^{2}}{4} |v_{s} - v_{p} |^{2},\nonumber
\\
m^{2}_{W^{(2)}} = \frac{g^{2}}{4} |v_{s} + v_{p} |^{2},\nonumber
\\
m^{2}_{Z^{(1)}} = m^{2}_{W^{(1)}}
\frac{2(g^{2}+g^{2}_{0})}{2g^{2}+g^{2}_{0}} \left[
1-\frac{g^{2}_{0}}{g^{2}} x +{\cal O}(x^{2})+{\cal
O}(g_{A}^{2})\right],\nonumber
\\
m^{2}_{Z^{(2)}} = m^{2}_{W^{(2)}}
\frac{(2g^{2}+g^{2}_{0})}{2g^{2}} \left[ 1+\frac{g^{2}_{0}}{g^{2}}
x +{\cal O}(x^{2})+{\cal O}(g_{A}^{2})\right],\nonumber
\end{gather*}
where $x= g^{2}g^{2}_{0}(2g^{2}+g^{2}_{0})^{-2}|v_{s}-v_{p}|^{2}
|v_{s}+v_{p}|^{-2}$.

The electroweak interaction Lagrangian is
\begin{gather}
{\cal L}_{\rm int} = e_{0} \overline{e} \gamma^{\mu} e A_{\mu} +
\frac{gg_{A}}{2\sqrt{2(g^{2}+(c_{2}g_{A})^{2})}} \left[
(c_{1}-c_{2}) \overline{\nu}_{e} \gamma^{\mu} \gamma^{5} \nu_{e} +
(c_{1}+c_{2}) \overline{e} \gamma^{\mu} \gamma^{5} e \right]
Y_{\mu}  \nonumber
\\
\phantom{{\cal L}_{\rm int} =}{}- \frac{g}{2\sqrt{2}} \left[
\overline{\nu}_{e} \gamma^{\mu} (1- \gamma^{5} ) e W^{(1)+}_{\mu}
+ \overline{\nu}_{e} \gamma^{\mu} (1+ \gamma^{5} ) e
W^{(2)+}_{\mu} + {\rm h.c.} \right]  \nonumber
\\
\phantom{{\cal L}_{\rm int} =}{}-\frac{g}{2}
\sqrt{\frac{2(g^{2}+g^{2}_{0})}{2g^{2}+g^{2}_{0}}} \Bigg[
\overline{\nu}_{e}\left( \frac{1}{2} \left( 1-
\frac{2(g^{2}+g^{2}_{0})} {g^{2}} x \right) \gamma^{\mu} -
\frac{1}{2} \left( 1+ 2x \right) \gamma^{\mu} \gamma^{5} \right)
\nu_{e}  + \nonumber
\\
\phantom{{\cal L}_{\rm int} =}{} + \overline{e} \left( \left(
-\frac{1}{2} + \frac{g^{2}_{0}} {g^{2}+g^{2}_{0}} \right) \left(
1- \frac{2(g^{2}+g^{2}_{0})}{g^{2}} x \right) \gamma^{\mu} +
\frac{1}{2} \left( 1+ 2x \right) \gamma^{\mu} \gamma^{5}
\right) e\nonumber\\
\phantom{{\cal L}_{\rm int} =}{} + {\cal O}(x^{2}) + {\cal
O}(g_{A}^{2})\Bigg] Z^{(1)}_{\mu} \nonumber
\\
\phantom{{\cal L}_{\rm int} =}{}-\frac{g}{2}
\sqrt{\frac{2g^{2}}{2g^{2}+g^{2}_{0}}} \Bigg[
\overline{\nu}_{e}\left( \left( \frac{1}{2} +
\frac{g^{2}_{0}}{2g^{2}} \right) \left( 1 + 2 x \right)
\gamma^{\mu} + \frac{1}{2} \left( 1 -
\frac{2(g^{2}+g^{2}_{0})}{g^{2}} x \right) \gamma^{\mu} \gamma^{5}
\right) \nu_{e}  \nonumber
\\
\phantom{{\cal L}_{\rm int} =}{} + \overline{e} \left( \left(
-\frac{1}{2} + \frac{g^{2}_{0}}{2g^{2}}
 \right) \left( 1 + 2 x \right)
\gamma^{\mu} - \frac{1}{2} \left( 1 -
\frac{2(g^{2}+g^{2}_{0})}{g^{2}} x \right) \gamma^{\mu} \gamma^{5}
\right) e  \nonumber
\\
\phantom{{\cal L}_{\rm int} =}{} + {\cal O}(x^{2}) + {\cal O}(g_{A}^{2}) \Bigg] Z^{(2)}_{\mu},\label{equation23}
\end{gather}
where $e_{0}= gg_{0}/\sqrt{2(g^{2}+g^{2}_{0})}$ is the positron
electric charge.

Any extension of some known chiral gauge model demands new
consideration of a question on the axial anomaly of
Adler--Bell--Jackiw \cite{Adler1,Bell}, which must be cancelled to
avoid the breakdown of gauge invariance and renormalizability of
the theory. The Adler--Bardeen theorem \cite{Adler2} guarantees
that the axial anomaly only receives contribution from one loop
diagrams, which are \cite{Adler3} the AVV or AAA triangle
diagrams. Freedom from the triangular chiral gauge anomaly was
first proved for the Weinberg--Salam model with $SU(2)_{L} \otimes
U(1)$  \cite{Gross,Bouchiat}, then for the standard model with
$SU(3)_{c} \otimes SU(2)_{L} \otimes U(1)$ and the left-right
symmetric model with $SU(3)_{c} \otimes SU(2)_{L} \otimes
SU(2)_{R} \otimes U(1)$ (see, for example~\cite{Geng}). It is
remarkable that in all mentioned cases, the anomaly exactly
cancels between leptons and quarks for each their family. The
anomaly-free conditions do not demand the existence of several
generations of fundamental fermions.

In our initially $P$-invariant electroweak model with light
(massless) axial gauge boson, the gauge group is locally
isomorphic to the group $SU(2)_{L} \otimes SU(2)_{R} \otimes U(1)
\otimes U(1)_{A}$. So, it is enough to consider the conditions of
the axial anomaly cancellation for triangular diagrams with one,
two, or three axial gauge bosons. It is easy to be convinced that
such cancellation will necessarily occur, at least, when the
constant $c_{2}$ from equations~\eqref{equation10},
\eqref{equation12}, \eqref{equation14}, and \eqref{equation23} is
equal to zero, and the constant $c_{1}$ from equations
\eqref{equation7}, \eqref{equation9}, and \eqref{equation23} is
equal to zero for one of the three lepton (quark) generations and
takes opposite values for the other two. Note that nonzero values
of the constant $|c_{1}|$ for leptons and quarks are not connected
in any way with each other.

\section{Existing models with a light or very light gauge boson}

A gauge boson, which may be very light and very weakly coupled,
was proposed a long time ago in the context of a supersymmetric
theory as the spin-1 partner of the goldstino (gravitino)
\cite{Fayet80,Fayet81}. It was supposed, that the new boson is
axially coupled to leptons and quarks.

A new light gauge boson introduced also by the consideration of an
extra $U(1)$ gauge symmetry for extension of the Weinberg--Salam
model or the left-right symmetric model \cite{Fayet90}. Let us
notice that in the work \cite{Fayet90} the extra gauge group is
$U(1)$, but is not $U(1)_{A}$. Therefore this gauge boson
interacts with the current, which is pure vector or it also
includes an axial part.

Some experimental consequences of the possible existence of a new
light gauge boson are discussed in the paper \cite{Boehm}. I would
like to add the following to it.

The fundamental question, it would be desirable to receive an
exhaustive answer to, concerns an old problem of the solar
neutrino (see, for example, the review \cite{Bellerive}). Can the
interaction of the light (massless) axial gauge boson give, as a
result, such energy spectrum of the solar neutrino which leads to
the reduction of the observable rate of transitions $^{37}{\rm
Cl}$ $\rightarrow$ $^{37}{\rm Ar}$ and $^{71}{\rm Ga}$
$\rightarrow$ $^{71}{\rm Ge}$ in comparison with the rate expected
in the standard solar model?

It is admissible to suppose that the light axial gauge boson
contribution to the elastic scattering of reactor antineutrinos on
electrons is comparable with the $Z$-boson contribution. This
statement is based on that the charged current gives the
dominating contribution in the cross section of
$\overline{\nu}_{e}e$ elastic scattering \cite{Marciano}, and the
experimental errors and background are large enough
\cite{Reines,Daraktchieva}. If $c_{1}= 2\sqrt{2}$ for the
electronic isodoublet, then the corresponding estimation of the
$\overline{\nu}_{e}e$ coupling constant gives the following upper
limit
\begin{gather*}
\alpha_{A} \equiv \frac{g_{A}^{2}}{4\pi} \sim 10^{-8} \alpha,
\end{gather*}
where $\alpha$ is the fine-structure constant. It is necessary to
emphasize especially that this estimation does not contradict to
the experimental value for the gyromagnetic quantity $g-2$ of the
electron~\cite{Particle}, having in mind the replacement of the
standard formula \cite{Levine} $g-2=\alpha/\pi+{\cal
O}(\alpha^{2})$ with the formula
$g-2=\alpha/\pi+\alpha_{A}/\pi+{\cal O}(\alpha^{2})+{\cal
O}(\alpha_{A}^{2})$. For the letter expression, the value of
$\alpha$ must be extracted from all available experimental data
\cite{Mohr}, except the ones regarding the $g-2$ factor itself.

\subsection*{Acknowledgements}

I am very grateful to S.P.~Baranov and N.P.~Yudin for useful
discussions on this report.

\LastPageEnding


\begin{thebibliography}{99}
\footnotesize

\bibitem{Slad00}
Slad L.M., Double symmetries in field theories, {\it Mod. Phys.
Lett. A}, 2000, V.15, 379--389, hep-th/0003107.

\bibitem{Gell-Mann}
Gell-Mann M., Levy M., The axial vector current in beta decay,
{\it Nuovo Cimento}, 1960, V.16, 705--726.

\bibitem{Slad01}
Slad L.M., Toward an infinite-component field theory with a double
symmetry: Free fields, {\it Theor. Math. Phys.}, 2001, V.129, 
1369--1384, hep-th/0111140.

\bibitem{Slad02}
Slad L.M., Toward an infinite-component field theory with a double
symmetry: Interaction of fields, {\it Theor. Math. Phys.}, 2002,
V.133, 1363--1375, hep-th/0210120.

\bibitem{Slad04}
Slad L.M., Double symmetry and infinite-component field theory, in
Proceedinds of Fifth International Conference ``Symmetry in
Nonlinear Mathematical Physics'' (June 23--29, 2003, Kyiv),
Editors A.G.~Nikitin, V.M.~Boyko, R.O.~Popovych, and
I.A.~Yehorchenko, {\it Proceedings of Institute of Mathematics},
Kyiv, 2004, V.50, Part~2, 947--954, hep-th/0312273.

\bibitem{Slad05}
Slad L.M., Mass spectra in the doubly symmetric theory of
infinite-component fields, {\it Theor. Math. Phys.}, 2005, V.142,
15--28, hep-th/0312150.

\bibitem{Ginzburg}
Ginzburg V.L., On relativistic wave equations with mass spectrum,
{\it Acta Physica Polonica}, 1956, V.15, 163--175.

\bibitem{Gelfand}
Gelfand I.M., Minlos R.A., Shapiro Z.Ya., Representations of the
rotation and Lorenz group and their applications, New York, The
Macmillan Company, 1963.

\bibitem{Pati}
Pati J.C., Salam A., Lepton number as the fourth ``color'', {\it
Phys. Rev. D}, 1974, V.10, 275--289.

\bibitem{Mohapatra1}
Mohapatra R.N., Pati J.C., Left-right gauge symmetry and an
``isoconjugate'' model of $CP$ violation, {\it Phys. Rev. D},
1975, V.11, 566--571.

\bibitem{Mohapatra2}
Mohapatra R.N., Pati J.C., ``Natural'' left-right symmetry, {\it
Phys. Rev. D}, 1975, V.11, 2558--2561.

\bibitem{Senjanovic}
Senjanovic G., Mohapatra R.N., Exact left-right symmetry and
spontaneous violation of parity, {\it Phys. Rev.~D}, 1975, V.12,
1502--1505.

\bibitem{Particle}
Eidelman S. et al., Review of particle physics, {\it Phys.
Lett.~B}, 2004, V.592, 1--1109.

\bibitem{Weinberg}
Weinberg S., A model of leptons, {\it Phys. Rev. Lett.}, 1967,
V.19, 1264--1266.

\bibitem{Adler1}
Adler S.L., Axial-vector vertex in spinor electrodynamics, {\it
Phys. Rev.}, 1969, V.177,  2426--2438.

\bibitem{Bell}
Bell J., Jackiw R., A PCAC puzzle: $\pi^{0} \rightarrow \gamma
\gamma$ in the sigma model, {\it Nuovo Cimento A}, 1969, V.60,
47--61.

\bibitem{Adler2}
Adler S.L., Bardeen W.A., Absence of higher-order corrections in
the anomalous axial-vector divergence equation, {\it Phys. Rev.},
1969, V.182, 1517--1536.

\bibitem{Adler3}
Adler S.L., Anomalies, hep-th/0411038.

\bibitem{Gross}
Gross D.J., Jackiw R., Effect of anomalies on quasi-renormalizable
theories, {\it Phys. Rev. D}, 1972, V.6, 477--493.

\bibitem{Bouchiat}
Bouchiat C., Iliopoulos J., Meyer Ph., An anomaly-free version of
Weinberg's model, {\it Phys. Lett. B}, 1972, V.38, 519--523.

\bibitem{Geng}
Geng C.Q., Marshak R.E., Uniqueness of quark and lepton
representations in the standard model from the anomalies
viewpoint, {\it Phys. Rev. D}, 1989, V.39, 693--696.

\bibitem{Fayet80}
Fayet P., Effects of the spin-1 partner of the goldstino
(gravitino) on neutral current phenomenology, {\it Phys. Lett.~B},
1980, V.95, 285--289.

\bibitem{Fayet81}
Fayet P., A la recherche d'un nouveau boson de spin un, {\it Nucl.
Phys. B}, 1981, V.187, 184--204.

\bibitem{Fayet90}
Fayet P., Extra $U(1)$'s and new forces, {\it Nucl. Phys. B},
1990, V.347, 743--768.

\bibitem{Boehm}
Boehm C., Implications of a new light gauge boson for neutrino
physics, {\it Phys. Rev. D}, 2004, V.70, 055007, 9 pages.

\bibitem{Bellerive}
Bellerive A., Review of solar neutrino experiments, {\it Int. J.
Mod. Phys. A}, 2004, V.19, 1167--1179.

\bibitem{Marciano}
Marciano W.J., Parsa Z., Neutrino-electron scattering theory, {\it
J. Phys.~G}, 2003, V.29, 2629--2646.

\bibitem{Reines}
Reines F., Curr H.S., Sobel H.W., Detection of
$\overline{\nu}_{e}-e$ scattering, {\it Phys. Rev. Lett.}, 1976,
V.37, 315--318.

\bibitem{Daraktchieva}
Daraktchieva Z. et al., Final results on the neutrino magnetic
moment from the MUNU experiment, {\it Phys. Lett. B}, 2005, V.615,
153--159.

\bibitem{Levine}
Levine M.J., Park H.Y., Roskies R.Z., High-precision evaluation of
contributions to $g-2$ of the electron in sixth order, {\it Phys.
Rev. D}, 1982, V.25, 2205--2207.

\bibitem{Mohr}
Mohr P.J., Taylor B.N., CODATA recommended values of the
fundamental physical constants: 2002, {\it Rev. Mod. Phys.}, 2005,
V.77, 1--107.

\end{thebibliography}
\end{document}